Defect properties of Na and K in $Cu_2ZnSnS_4$ from hybrid functional calculation

Kinfai Tse[1], Manhoi Wong[1], Yiou Zhang[1], Jingzhao Zhang[1], Michael Scarpulla[2], Junyi Zhu[1,*]

[1] Physics Department, The Chinese University of Hong Kong, Shatin, NT, Hong Kong SAR, The People's Republic of China

[2] Department of Materials Science and Engineering, University of Utah, Salt Lake City, UT 84112, USA

In-growth or post-deposition treatment of $Cu_2ZnSnS_4$ (CZTS) absorber layer had led to improved photovoltaic efficiency, however, the underlying physical mechanism of such improvements are less studied. In this study, the thermodynamics of Na and K related defects in CZTS are investigated from first principle approach using hybrid functional, with chemical potential of Na and K established from various phases of their polysulphides. Both Na and K predominantly substitute on Cu sites similar to their behavior in $Cu(In,Ga)Se_2$, in contrast to previous results using the generalized gradient approximation (GGA). All substitutional and interstitial defects are shown to be either shallow levels or highly energetically unfavorable. Defect complexing between Na and abundant intrinsic defects did not show possibility of significant incorporation enhancement or introducing deep n-type levels. The possible benefit of Na incorporation on enhancing photovoltaic efficiency is discussed. The negligible defect solubility of K in CZTS also suggests possible surfactant candidate.

PACS numbers(s): 61.72.-y, 71.55.-i, 84.60.Jt

---

[*] Corresponding author, jyzhu@phy.cuhk.edu.hk

I.   INTRODUCTION

Cu$_2$ZnSnS$_4$/Cu$_2$ZnSnSe$_4$ (CZTSSe alloy system) is a promising photovoltaic absorber material composed of non-toxic, earth abundant elements, having a tunable band gap between 1.0 eV for CZTSe and 1.5 eV at CZTS limit and a high absorption coefficient.[1] The record efficiency of CZTSSe devices reached 12.6% as of 2014,[2] yet increasing this efficiency and especially the open circuit voltage Voc has remained a challenge. Increasing the doping in CZTSSe in part by lowering the activation energy of acceptors could help to address the Voc challenge. Also, in the close cousin Cu(In,Ga)Se$_2$ (CIGSe) the presence of Na during absorber layer growth[3] as well as group I post deposition treatments of the absorber layer interface[4–7] have helped to maximize solar cell efficiency.

Experiments have established that incorporation of Na in CZTS/CZTSe growth would results in improvements in electrical properties[8–13] and morphology.[11,12] However, excess Na has also been demonstrated to reduce device performance.[11] It is therefore important to assess the thermodynamics of group I doping and alloying in CZTS and the corresponding changes in electronic states induced. In chalcopyrite CIGSe, Na primarily substitutes for Cu$^{1+}$ ions[14] and it is speculated that Na shows similar defect properties in CZTSSe. While KF post deposition treatment has been well documented for CIGSe, it has been studied to a lesser extent in CZTSSe.[15] Therefore, establishing the thermodynamic limit of Na and K in CZTS from first principle calculation can serve as a first step to understand the underlying physical mechanism of the enhanced electrical properties.

Recently, there have been investigations on Na/K defects in CZTSe using GGA-PBE functional[16] and HSE hybrid functional.[17] Apart from the inadequacy of GGA functional as pointed out by Ref. 17, detailed investigations of the chemical potentials and phase diagrams have been lacking. Selecting a chemical potential outside of the CZTSSe single phase region places the system into an unphysical thermodynamic situation. Moreover, hybrid functional result of Na or K doping on CZTS is lacking, such result would provide better understanding of the defect properties in CZTS, which is considerably different from that of CZTSe. Knowledge of defect complex formation of Na and K with abundant intrinsic defects is important in identifying surfactant candidates, as defect complex formation can result in charge and stress compensation[18] which drastically lower the enthalpy of formation, increasing solubility. In this paper, based on first principles density functional theory calculations, the thermodynamic properties of Na and K point defect and defect complexes are investigated and expected concentrations of Na and K in CZTS are estimated.

## II. COMPUTATIONAL METHOD

Our calculations follow the method introduced in reference.[18] The formation energy of defect D in charge state q can be expressed as

$$\Delta H_f(D, q) = (E(D, q) - E_{host}) + \sum n_i(E_i + \mu_i) + qE_F \qquad (1)$$

where the 1st term is the difference of energy between defect and host supercell, the 2nd term is the free energy gain from removing $n_i$ atoms of species i from the perfect supercell and moved to the reservoir of chemical potential $\mu_i$ to form the defect supercell, and the 3rd term is the energy gained by removing q electrons from the supercell to an electron reservoir having Fermi energy $E_f$. The transition energy level between charge states q and q' of the same defect is then defined as the Fermi energy at which their formation energies are equal. The chemical potentials for atoms i can be varied within the CZTS single phase region bounded by competing second phases established by the requirement that formation of secondary phases are energetically unfavorable.[19] To avoid segregation of elements, the computed $\mu_i$ is always negative, thus the chemical potential acts as an extra energy penalty to addition of an atom and extra energy gain from removal of an atom. The determination of the chemical potentials for group I elements such as Na and K is particularly complicated by the fact that they form polysulphides of various chain length. For each of the polysulphides, it is possible to solve for the richest $\mu_{Na}$ achievable given the formation energy $\Delta H_f(Na_xS_y)$ and the sulphur chemical potential $\mu_S$ from the inequality

$$\Delta H_f(Na_xS_y) = E(Na_xS_y) - x(E_{Na} + \mu_{Na}) - y(E_s + \mu_s) \geq 0 \qquad (2)$$

To satisfy the constraint simultaneously for all polysulphide phases means to take the most constraining chemical potential

$$\mu_{Na} = min(0, \mu_{Na}(Na_2S), \mu_{Na}(NaS), \mu(Na_2S_4), \mu_{Na}(Na_2S_5)) \qquad (3)$$

In this work, pure metallic elements, sulphur molecules[20,21] with chemical formula $S_2$, $S_6$ and $S_8$, binary and ternary secondary phases as listed in related works[19,22,23] were considered in the construction of CZTS phase diagram. The chemical potentials of Na and K were determined from polysulphides $(Na/K)_xS_y$, with the ratio of x:y ranging from x:y = 2:1 to 2:6.[24,25] Within the quasichemical formalism for point defects, the concentrations of each charge state of each type of defect may be determined from

$$N = N_{site} \cdot g \cdot e^{-H_f(D,q)/k_BT} \qquad (4)$$

Thus, the system of M equations of form of equation (4) plus the charge balance equation can be solved self consistently for the defect concentrations and Fermi level. Herein we calculated the Fermi energy at a typical growth temperature of 900 K.[9,10]

Point defects are modelled by the introduction of single defect in periodically placed supercells constitute of 64 atoms, with reciprocal space sampled with 2×2×2 Γ-centered k-points sampling scheme in the first Brillouin zone. For defect complexes formed by 2 defects, a supercell of 128 atoms is used. Total energy calculations were performed using the projected augmented-wave method with the HSE[26] functional as implemented in the Vienna *ab initio* simulation package.[27] The transition levels are calculated using the special k-point scheme[18] and accounting for core level shifts. A plane-wave basis set with 450 eV cutoff was used, the accuracy in total energy was set at $10^{-6}$ eV per atom and the internal structure of host and defect supercell were relaxed until forces were below 0.01 eV/Å.

TABLE I. Numerical value of chemical potential at the chosen points of interest.

| Δμ (eV) | Cu | Zn | Sn | S | Na | K |
|---|---|---|---|---|---|---|
| A | -0.74 | -1.90 | -1.39 | 0 | -1.94 | -2.23 |
| B | -0.51 | -2.05 | -1.68 | 0 | -1.94 | -2.23 |
| C | -0.44 | -1.30 | -0.20 | -0.60 | -1.46 | -1.37 |
| D | -0.24 | -1.10 | 0 | -0.80 | -1.36 | -1.27 |
| E | 0 | -1.06 | -0.36 | -0.83 | -1.34 | -1.25 |

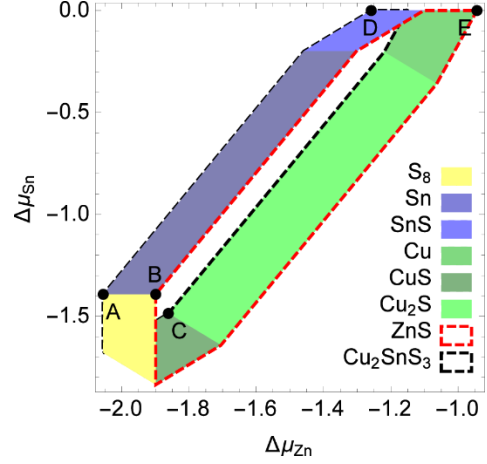

FIG. 1. Allowed chemical potential range of Cu2ZnSnS4 calculated with hybrid functional, points correspond to (A) Cu-poor, (B) lowest and (D) highest Fermi-energy level, (C) lowest and (E) highest Na incorporation.

## III. RESULTS AND DISCUSSION

Fig. 1 shows the phase diagram of CZTS, the corresponding chemical potentials are tabulated in Table 1. The calculated phase diagram for intrinsic elements is in agreement with previous work.[23] Moreover, inclusion of sulphur-rich polysulphide phases impose further restriction on the chemical potential $\mu_{Na}$ and $\mu_K$. Compared to results obtained from Na and K elemental phase, the chemical potentials are lowered by 1.76 eV and 1.67 eV respectively. With respect to considering $Na_2S$ or $K_2S$, the chemical potentials were still lowered by 0.18 eV and 0.56 eV, which amounts to 1-3 order of magnitude difference in defect concentration.

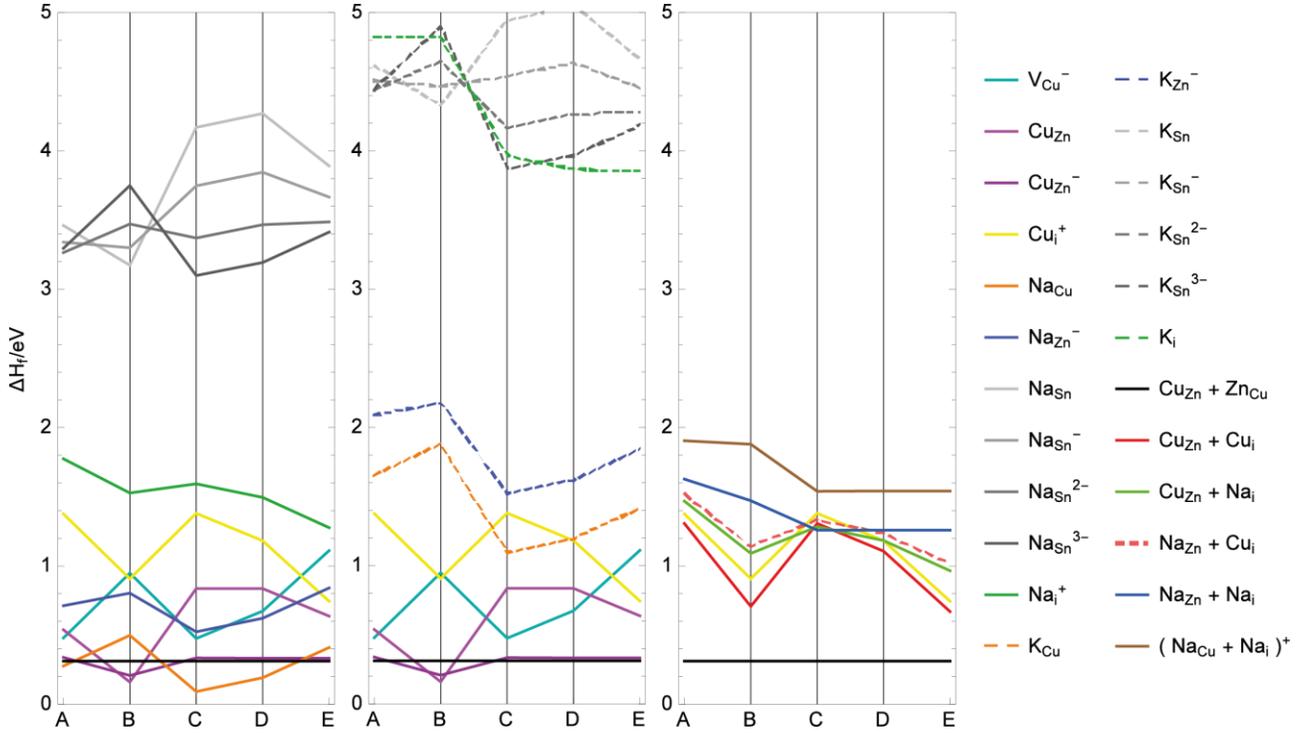

FIG. 2. Formation energy of Na (left), K (middle) point defect, and defect complexes (right), including representative intrinsic defects.

The formation energy of Na and K point defects and defect complexes are plotted in Fig. 2. The formation energy of $V_{Cu}$, $Cu_{Zn}$ and $Cu_i$ intrinsic defects are shown for comparison, calculated intrinsic defects also show good agreement with previous work.[22,23] We considered Na or K substitution on either of the two crystallographically inequivalent Cu sites, Zn or Sn sites, and 3 interstitial sites as shown in Fig. 3, two of which are octahedral coordinated (oc) at the Sn and Zn layer respectively, and one of which is tetrahedral coordinated (tc) with 4 metallic ions. The formation energy of K defects is generally >1eV higher than the Na counterparts, due primarily to the larger atomic size of K. Of the substitutional defects, $Na_{Cu}$ in either sites, with energy difference only in meV range, are the most thermodynamically stable defect over all chemical potentials within the single phase kesterite region, followed by $Na_{Zn}$. The formation of $Na_{Sn}$ substitutional defects is unlikely as its formation energy is at least 2.9 eV greater than the formation energy of $Na_i$ for all chemical potentials considered. Amongst the 3 inequivalent interstitial sites, $Na_i$ prefers the tc site, while $K_i$ prefers the oc site with 2 Zn second-nearest neighbor, however the formation of $K_i$ is also highly unlikely.

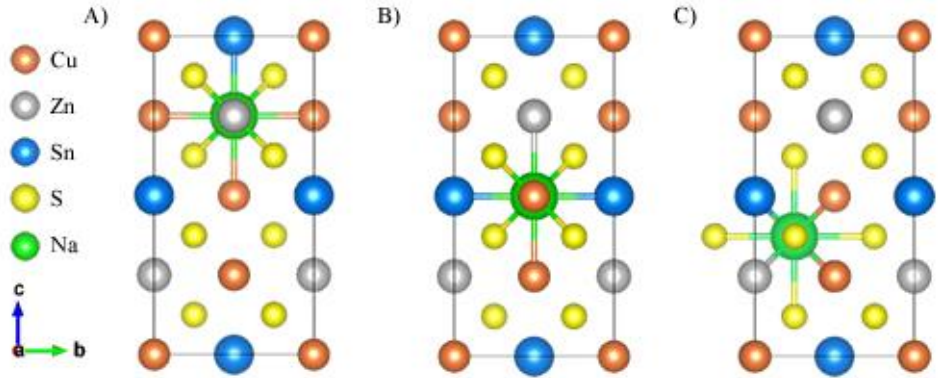

FIG. 3. Inequivalent interstitial sites (A) Zn layer oc, (B) Sn layer oc, and (C) tc site with metallic ions, before relaxation, illustrated in a unit cell.

The formation energies of $Na_{Zn}$ and $K_{Zn}$ calculated with the HSE hybrid functional are found to be significantly higher than $Na_{Cu}$ and $K_{Cu}$ over a wide range of chemical potential, in contrast with previously reported results.[16] Similar over-estimation of stability and negative formation energy can also be seen in $Cu_{Zn}$ when comparing results from GGA[19] to that of HSE.[22,23] A comparison with our GGA-PBE calculation shows that PBE severely overestimate the stability of p-type defects instead of the commonly believed n-type defect, possibly due to overestimation of Cu-3d S-2p coupling by the GGA functional,[28,29] resulting in an artificially high valence band edge. Indeed, a rigid shifting of valence band instead would result in decent qualitative agreement over most of the intrinsic p-type defects, Na and K substitutional defects.

Table 2 shows the computed the formation energies at point A for defect complexes consisting of a single interstitial with a single acceptor. Compared with isolated $Na_i^+$, the formation energy of such complexes is significantly reduced due to Coulombic attraction between the ionized donor and a single acceptor such as $Na_{Zn}$ or $Cu_{Zn}$ which both have relatively high concentrations in CZTS. $Na_{Cu}+Na_i$ is energetically less favorable over the whole chemical potential range compared to other defect complexes, mainly due to the absence of this Coulombic attraction. As shown in Table 2, due to the low electronegativity of Na, $Na_{Zn}+Na_i$ has the strongest interaction. However, the stability is also determined by the energy of the isolated defects in the limit of infinite separation. Summing both contributions results in $Cu_{Zn}+Na_i$ and $Na_{Zn}+Cu_i$ being the most stable defect complexes while the formation energy of $Na_{Zn}+Na_i$ is slightly higher. This lowering of formation energy through defect complexing did not result in a significant incorporation enhancement for Na as all the formation energies all exceed 1.47 eV.

TABLE II Formation energy of defect complex compared to separated point defects.

| ΔH (eV) | $Na_{Zn}+Na_i$ | $Cu_{Zn}+Na_i$ | $Na_{Zn}+Cu_i$ | $Cu_{Zn}+Cu_i$ | $(Na_{Cu}+Na_i)^+$ |
|---|---|---|---|---|---|
| ΔH Separated | 2.49 | 2.11 | 2.09 | 1.71 | 2.05 |
| ΔH Complex | 1.63 | 1.47 | 1.52 | 1.31 | 1.90 |
| E Interaction | 0.86 | 0.64 | 0.57 | 0.41 | 0.15 |

The formation energies of the most stable charge states of defects as a function of Fermi energy level are shown in Fig. 4. Chemical potential at point B represents the Cu-poor, Zn-poorest and Sn-poor condition, where the formation of the major n-type antisite defect $Sn_{Zn}$ and $Zn_{Cu}$ is unfavorable, resulting in the lowest Fermi-energy level. Chemical potential at point D represents a Sn-richest, Zn-rich and Cu-mild condition where the formation energies of most n-type defects are lowered, and hence the highest allowable Fermi-energy level in equilibrium close to the mid gap. Both $Na_{Zn}$ and $K_{Zn}$ display a shallower (0/1-) transition energy level than $Cu_{Zn}$. Our calculation on intrinsic point defect shows qualitative agreement with previous work,[22,23] the calculated $V_{Cu}$ (0/-1) transition is 7 meV below the VBM, and Cu$_{Zn}$ (0/-1) transition is at 0.23 eV above the VBM. Over the range of possible self-consistently determined Fermi energies, Na$_{Zn}$ and K$_{Zn}$ are almost always ionized to the -1 charge state, while Na$_{Sn}$ and K$_{Sn}$ can vary from 0 to -3 in different growth condition. Ionization of Na$_{Zn}$ is possibly one source of contribution to the observed increase in p-type conductivity.[8,12,30] The transition level of Na$_i$ is not within the bandgap, and a charge state of +1 is always highly preferred. The formation of defect complex between Na$_{Cu}$ and Na$_i$ induces a shift of transition level of Na$_i$ to 0.12 eV below the conduction band minimum as a shallow donor.

Because of the low formation energy, the highest concentration of Na$_{Cu}$ is achieved at point B with about 2 order of magnitude difference between concentration of Na$_{Cu}$ and Na$_{Zn}$, showing that although Na can effectively act as a p-type dopant, Na predominantly act as an isovalent substituent of Cu in CZTS, similar to CIGS.[9] The concentration of K defects is generally 8 orders of magnitudes lower than

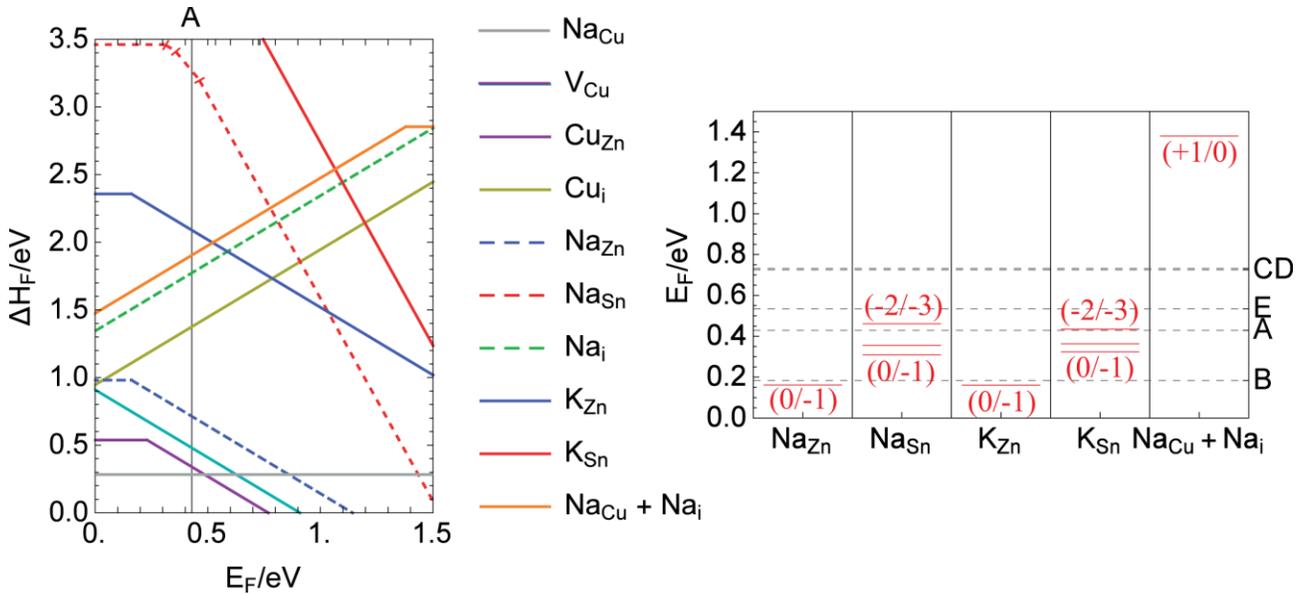

FIG. 4. Formation energy of defects in the most stable charge state for Na/K and selected intrinsic defects at chemical potential point A (left) and defect transition energies within the band gap (right). The Fermi energy is shown as gray lines labelled for the different chemical potential points.

Na defects, this low defect solubility combined with the benign electronic properties of the defects may suggest that K is a good surfactant candidate.[15] Lattice expansion has been observed in Na-doped single crystalline CZTS.[8,30] Our calculations show that both $Na_{Cu}$ and $Na_{Zn}^-$ would result in a lattice expansion, therefore the observed lattice expansion is probably caused by $Na_{Cu}$ or $Na_{Zn}$ doping. The band gap of CZTS is widened by 67 meV upon single $Na_{Cu}$ substitution in the supercell, due to reduced Cu-3d S-2p coupling, thus introducing Na at the buffer-absorber interface could in principle contribute to enhanced open-circuit voltage.

## IV. CONCLUSION

Using HSE hybrid functional first principle calculations, we studied the point defects associated with Na and K, and established the thermodynamic limit of Na and K in CZTS. We determined that Na and K polysulphides impose more stringent restrictions on the boundaries of the kesterite single-phase region in the chemical potential space. The lowest energy configuration of Na and K defects is isovalent substitution on Cu sites to form $Na_{Cu}$ or $K_{Cu}$, in contrast to previous reports. Formation of any K defects is unlikely compared to Na yet if K was incorporated it would result in shallow acceptor levels and hence might increase photovoltaic performance. $Na_{Zn}$ is a shallow acceptor with transition level shallower than $Cu_{Zn}$. Although $Na_{Sn}$ form relatively deep acceptor levels, similar to most Sn-related defects in CZTS, it is not problematic as the concentration will be exceptionally low. Complex formation of interstitials with single acceptor $Na_{Zn}$ or $Cu_{Zn}$ leads to a decreased energy for interstitial defect formation, however Na still predominantly resides in Cu or Zn antisites. This knowledge contributes towards the understanding of possible surfactant or passivation effect of Na and K in the grain boundary.

## ACKNOWLEDGEMENTS

Financial support of General Research Fund (2130490) and Research Incentive Scheme (4441641) from the Research Grants Council in Hong Kong is gratefully acknowledged.